\documentclass[sigconf]{acmart}

\setcopyright{none}
\acmDOI{}
\acmISBN{}

\acmConference[LIMITS '26]{12th Workshop on Computing within Limits}{June 23--25, 2026}{Online}

\renewcommand\footnotetextcopyrightpermission[1]{}

\usepackage{graphicx}
\usepackage{microtype}
\usepackage{placeins}
\begin{document}

\title{Dismantle and Dissolve, (Re)build, Remix: A Research-creation Inquiry into the Political Economy of Graphics Cards}

\author{Cyrus Khalatbari}
\affiliation{%
  \institution{École de technologie supérieure (ÉTS)}
  \city{Montreal}
  \country{Canada}
}
\affiliation{%
  \institution{HEAD – Genève (HES-SO)}
  \city{Geneva}
  \country{Switzerland}
}
\email{cyrus.khalatbari@etsmtl.ca}

\begin{abstract}
This contribution follows a four-year investigation (2022--2026) into the political economy of graphics card miniaturization. It begins from the premise that rethinking our relationship to artificial intelligence and its sociotechnical entanglements requires demystifying and opening the black box of this technical object. Within our algorithmic culture, the graphics card (GPU) enables the massive, parallel processing of large datasets, making possible the training of the models that underpin our intelligent systems. GPU miniaturization is equally crucial: as a key driver of the Internet of Things, this sociotechnical phenomenon enables the inclusion of these cards in increasingly compact and powerful systems while also enabling better management of energy resources. The development of these everyday objects and technologies nevertheless reinforces several major problems. Drawing on both the social sciences and the critical, reflexive, speculative, and fictional methodologies of research-creation, the author developed several investigative fieldwork sites---among liquid nitrogen overclockers in Taiwan and urban miners in Ghana---and conducted situated experimentations on some fifty acquired graphics cards. Structured around three themes (dismantle and dissolve, rebuild, remix), this paper demonstrates how research-creation methods constitute full epistemologies for apprehending what seems a priori external, opaque, or inaccessible, and for restoring artificial intelligence to its tangible materialities. In doing so, it contributes to the field of ICT for sustainability by affirming research-creation as a rigorous means of disentangling the material and environmental infrastructures that computational systems both depend on and obscure.
\end{abstract}



\maketitle

\section{Miniaturization and its Stakes: General Context}

This contribution follows a four-year investigation (2022--2026) I conducted into the political economy of graphics card miniaturization\footnote{Doctorate completed jointly between the École Polytechnique Fédérale de Lausanne (EPFL) and HEAD – Geneva (HES-SO), University of Art and Design, Switzerland.}. More specifically, this article focuses on the situated experimentations I conducted with three collaborators on subsets of the fifty graphics cards acquired during this investigation. It begins from the premise that rethinking our relationship to artificial intelligence and its sociotechnical entanglements requires demystifying and opening the black box \cite{latour1999}\footnote{According to Latour, the black-boxing of technology occurs when "scientific and technical work is made invisible by its own success" (p. 304), rendering it opaque, intangible, and difficult to understand for its users.} of this technical object. Within our algorithmic culture, this component of our computers is critical. By virtue of its semiconductor architecture, the graphics card (GPU) enables the massive, parallel processing of large datasets, making possible the training of the models that underpin our intelligent systems. At the heart of the imperative toward connectivity and relentless performance on which this industry rests, GPU miniaturization is equally crucial. As a key driver of the Internet of Things, this sociotechnical phenomenon enables the inclusion of these cards in increasingly compact and powerful systems, while also enabling better management of energy resources. The development of these everyday objects and technologies nevertheless reinforces several major problems, which in turn affect our capacity to apprehend them and situate them within the broader contexts that structure them. Intimately linked to miniaturization, one of these problems lies in their compression and their very small scale, which prevents us from actually perceiving the imbrication of the various technical schemes\cite{simondon1958}\footnote{Simondon's technical schemes refer to the fundamental operative structures that allow us to understand how technical objects function. These are the organizational principles and functional relationships underlying a technical object, making intelligible the way its various components interact and cooperate to accomplish a given function.} and components that constitute them.

As demonstrated in my doctoral research, this material obscuration impacts three key characteristics that would otherwise allow us to better understand these objects and thus continue demystifying artificial intelligence. First, these engineering and design choices prevent us from perceiving and understanding the geological \cite{parikka2015} and elemental \cite{peters2015} ramifications of these artifacts---in other words, the various metals and energy requirements that enable the creation of these objects and the training of our algorithms within them. This dynamic of opacity produces a second dangerous slippage, defined by a decontextualization of these technologies from the ensemble of cultural, social, and labor practices that shape them. Finally, by rendering invisible the situated knowledges\cite{haraway1988} \footnote{For Haraway, situated knowledges refer to the idea that all knowledge is produced from a particular, embodied, and partial position. No knowledge can claim universal objectivity or a neutral viewpoint "from nowhere": every researcher occupies a specific position that shapes what they can see and understand.} articulated during the production and maintenance of graphics cards, miniaturization negatively impacts our understanding of the relations of power and control it engenders. It obscures in particular the asymmetric relations between corporations and communities of workers present throughout the chain of production, optimization, and recycling of these objects. This work thereby contributes to the field of ICT for sustainability, further reinforcing research-creation as a productive epistemological stance for disentangling the material and environmental realities that computational systems tend to render invisible. In doing so, these methods prove equally key in making visible the seams \cite{ratto2007} of these infrastructures and extend, in their own ways, the infrastructural inversion proposed by Bowker \cite{bowker1994}: turning these objects inside out to reveal the social, political, and material arrangements they silently stabilize.

\section{How I Conducted This Investigation}

\begin{figure}[!htbp]
  \centering
  \includegraphics[width=\columnwidth]{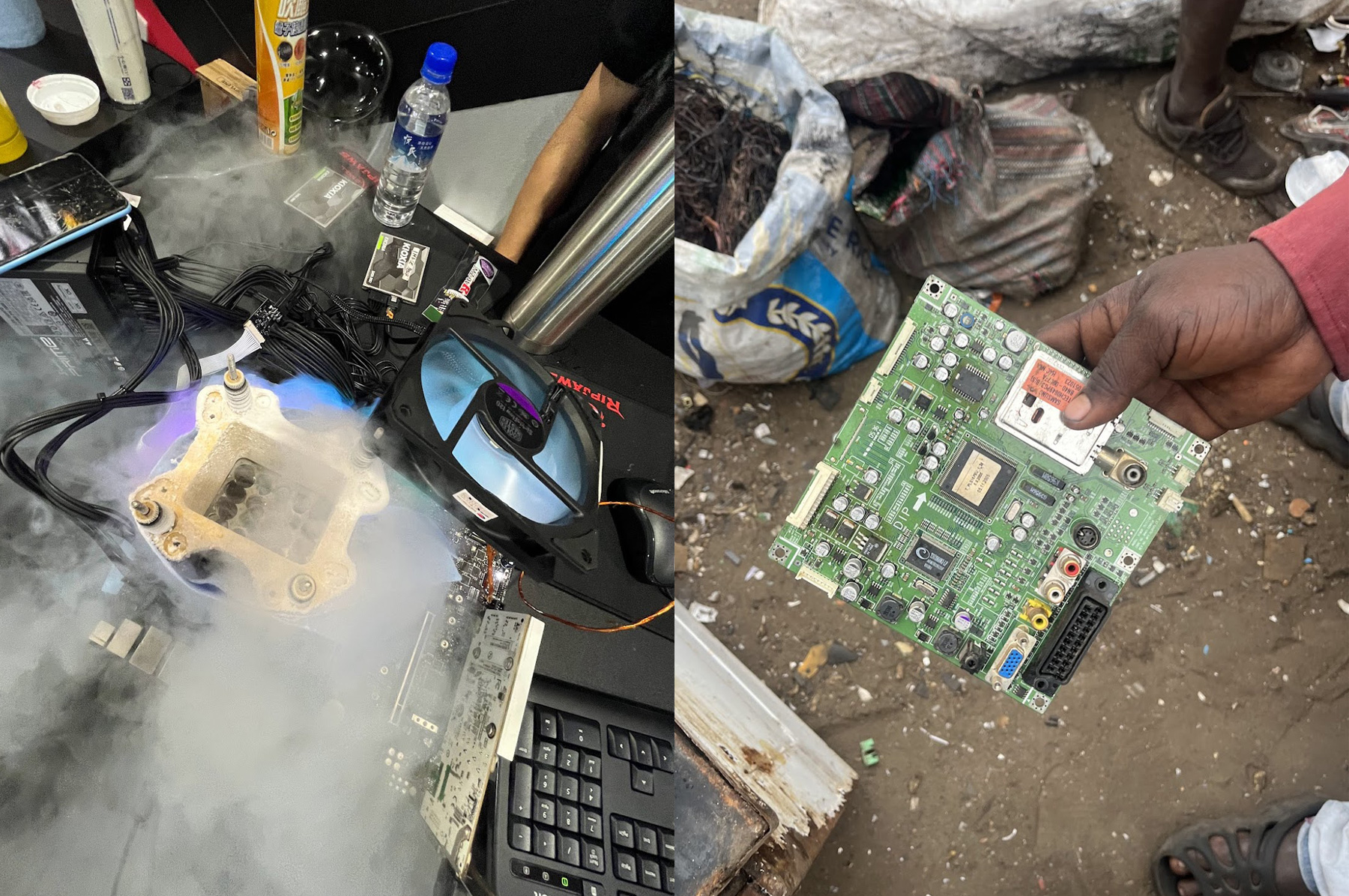}
  \caption{Liquid nitrogen overclocking (Taiwan); Urban mining (Ghana). Credit and source: Cyrus Khalatbari.}
  \label{fig:fieldwork}
\end{figure}

Drawing on both the social sciences and the critical and reflexive \cite{sengers2005}, speculative and fictional \cite{bleecker2009} methodologies of research-creation\footnote{As defined by the Hexagram academic network, research-creation links the interpretive disciplines with creative ones, involving "the creation of knowledge in and through creative material and performative practice" \cite{hexagram2025}.}, I developed several investigative fieldwork sites over the course of the past four years. To globalize these issues and better understand the environmental, social, and political ramifications of graphics card miniaturization, I first conducted two complementary investigations. The first led me to liquid nitrogen overclockers \footnote{Overclocking consists of running a component (CPU, GPU, RAM) at a higher frequency than that intended by the manufacturer in order to achieve better performance, which requires more power and consequently generates greater heat.} gathered in Taiwan at events like Computex, where computing power is pushed to its limits through spectacular extreme cooling practices [Fig.~\ref{fig:fieldwork}]. The second unfolded among urban miners in the Circle and Agbogbloshie neighborhoods of Accra, who dismantle, repair, and recycle these same components within informal circuits of remarkable inventiveness [Fig.~\ref{fig:fieldwork}]. Both investigations were grounded in sustained ethnographic engagement: semi-structured interviews with practitioners, systematic photographic documentation, and a research diary maintained throughout, recording observations, material encounters, and emerging analytical threads. Together, these sites offered a unique situated vantage point on the full arc of these objects---from peak performance to material afterlife---illuminating how communities of optimization and maintenance, across the global North and South, bear witness to the environmental, cultural, and labor consequences that miniaturization tends to obscure. Through these two fieldwork sites, I was able to reveal how the operational sequence \cite{leroi-gourhan1964} of these cards traverses radically different worlds, while being structured by the same logics of miniaturization, privatization, and an ever more exponential corporate control over communities of use. Drawing from these fieldwork sites, it is then through research-creation methods that I extended my research: further exposing, in situated ways, the political economy of computing power from the inside out.

\begin{figure}[!htbp]
  \centering
  \includegraphics[width=\columnwidth]{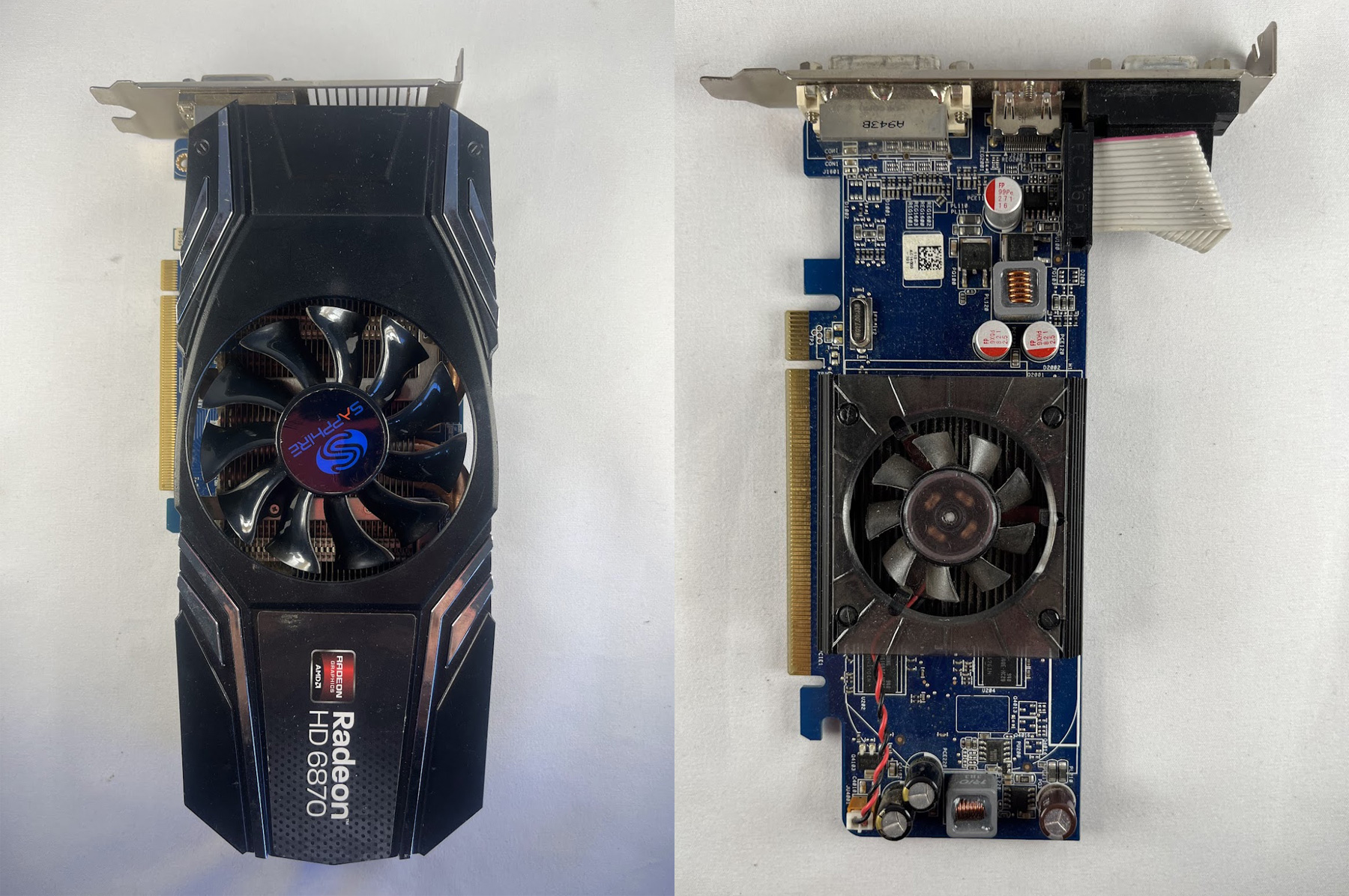}
  \caption{Two acquired graphics cards. Credit and source: Cyrus Khalatbari.}
  \label{fig:cards}
\end{figure}

My approach was also doubled by situated experimentations, carried out on some fifty graphics cards acquired in Ghana, Taiwan, and online [Fig.~\ref{fig:cards}]. Positioning myself as a research-through-design \cite{findeli2000} practictionner, it seemed essential to actively and collectively rethink the gestures observed in the field. Through three collaborations spread over a year, I condensed these experiences into a ninety-page field notebook. These collaborations involved three interlocutors with complementary profiles---a chemist and electronics specialist, a software engineer passionate about gaming, and a self-taught tinkerer specializing in electronic components---whose role was decisive: without their technical competencies, several entire sections of the research-through-making would simply not have been accessible. I will introduce them in more detail in each of the sections that follow.

\section{Demystifying through Making: Three Propositions}

In this text, I propose to condense several explorations conducted in this field notebook, structured around three complementary themes and ways of critically rethinking our graphics cards through making. Each section is organized, first, around the raw account of these experimentations, guiding the reader through our iterations and attempts. Drawing from these situated explorations and in dialogue with my fieldwork in Ghana and Taiwan, I then trace broader correlations around computing power and its environmental, social, and material implications---dimensions that remain buried inside this black box. The first section is defined by the practices of dismantling \cite{gaboury2018} and dissolution. By removing, weighing, grinding, and chemically transforming the various components of these technologies---chassis, fans, heat sinks, circuit boards, capacitors, and scattered components---I was able to retrace the geological and elemental ramifications of these artifacts, so often concealed by their manufacturers. My second theme is reconstruction \cite{ratto2011}. Through this practice, I reproduced gestures and artifacts observed in my fieldwork in order to better delineate their sociocultural and use contexts, necessary for their understanding and critical analysis \cite{nolan2018}. Finally, my third orientation anchors itself around a practice of remixing and tinkering. This approach, constituted around the production of hybrid and subversive \cite{certeau1980} objects combining deconstruction and reconstruction, allowed me to speculate and propose new lines of inquiry, collectively opening toward more sustainable sociotechnical imaginaries.

\subsection{27 March, 4 April, 5 May 2025: Dismantle and Dissolve}

\begin{figure}[!htbp]
  \centering
  \includegraphics[width=\columnwidth]{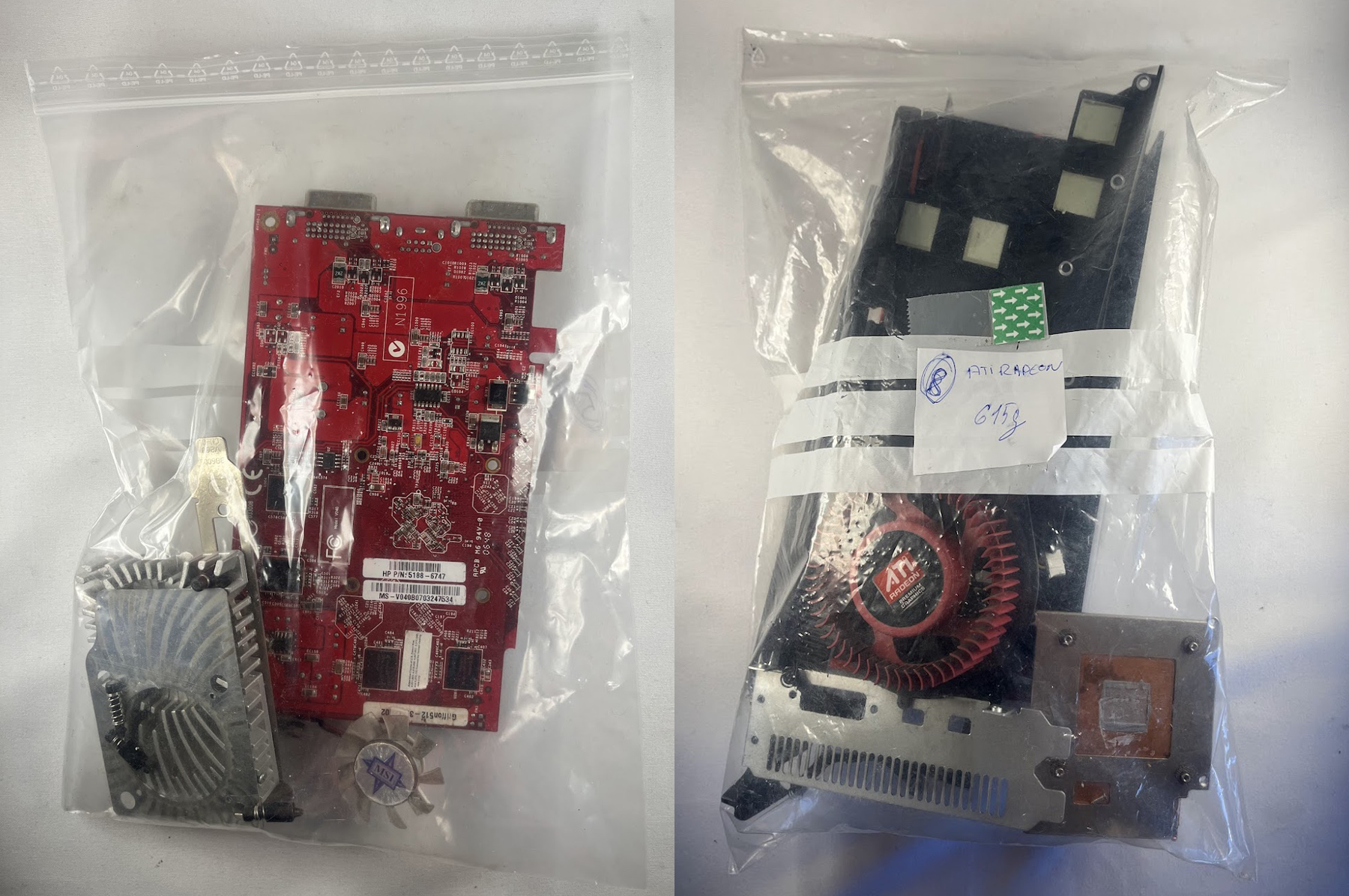}
  \caption{Disassembled graphics cards stored in Ziplocs. Credit and source: Cyrus Khalatbari.}
  \label{fig:ziploc}
\end{figure}

\begin{figure}[!htbp]
  \centering
  \includegraphics[width=\columnwidth]{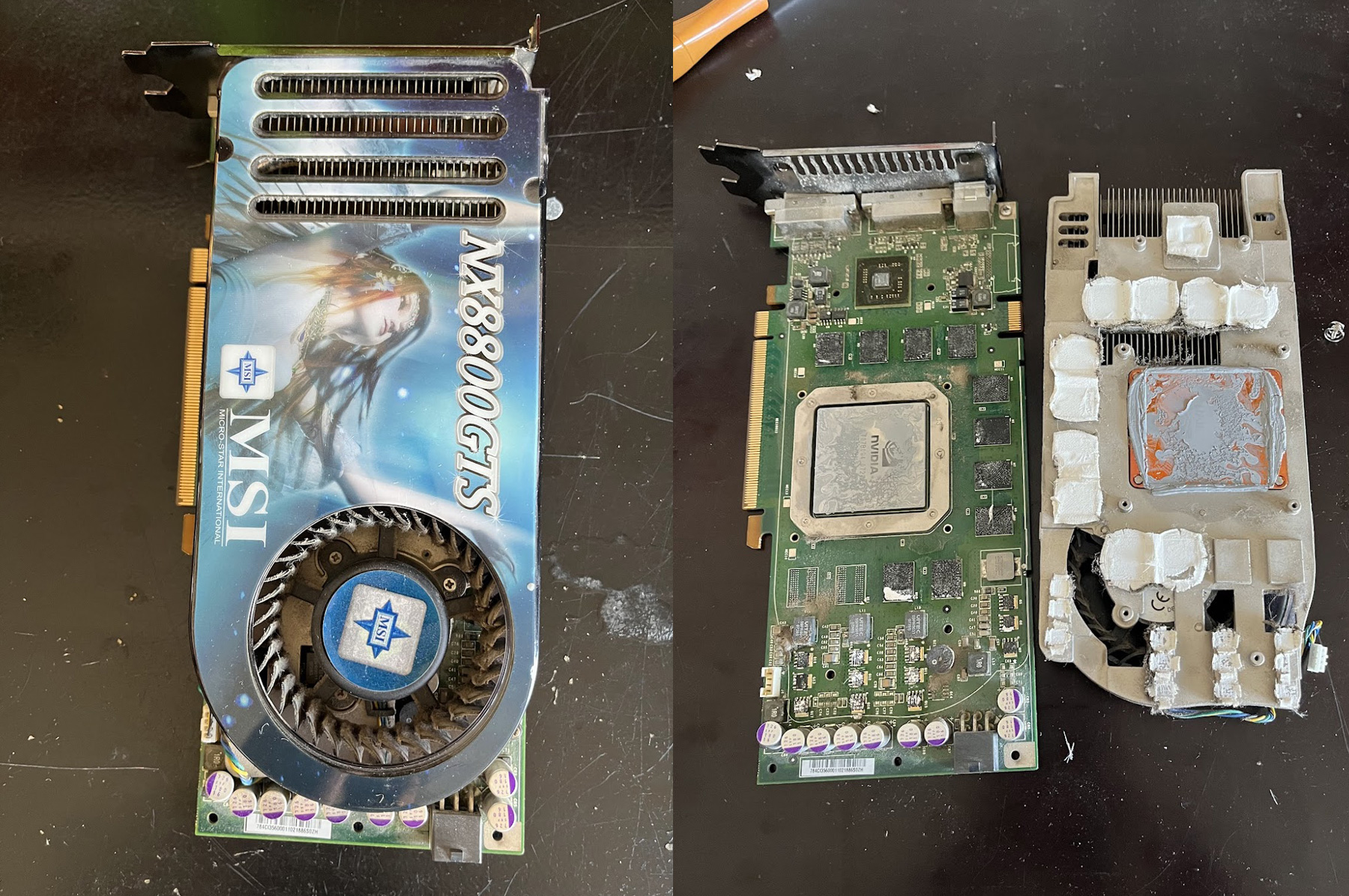}
  \caption{Graphics card with its circuit board (right). Credit and source: Cyrus Khalatbari.}
  \label{fig:circuitboard}
\end{figure}

\begin{figure}[!htbp]
  \centering
  \includegraphics[width=\columnwidth]{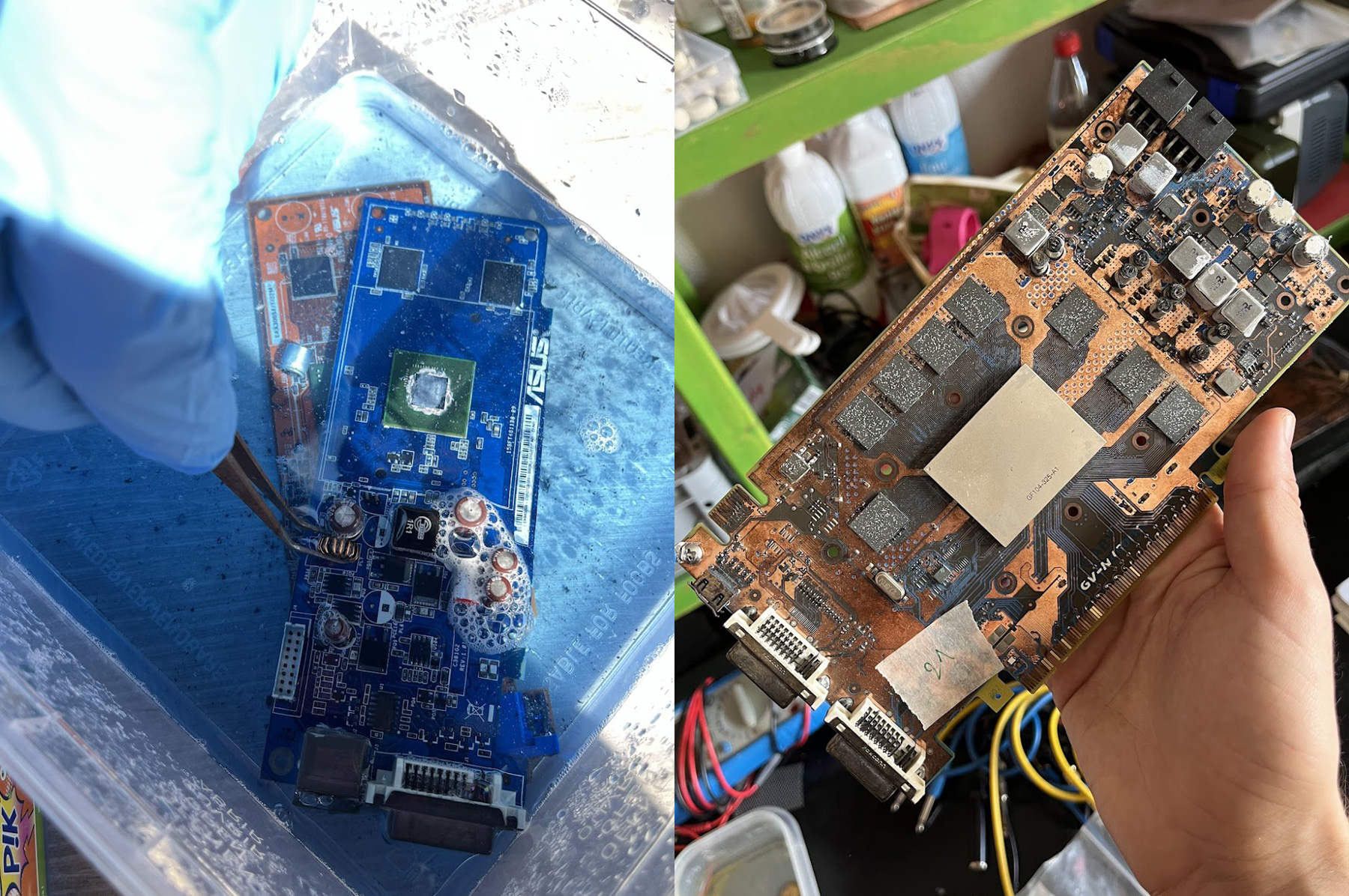}
  \caption{Use of caustic soda to dissolve the varnish. Credit and source: Cyrus Khalatbari.}
  \label{fig:caustic}
\end{figure}

\begin{figure}[!htbp]
  \centering
  \includegraphics[width=\columnwidth]{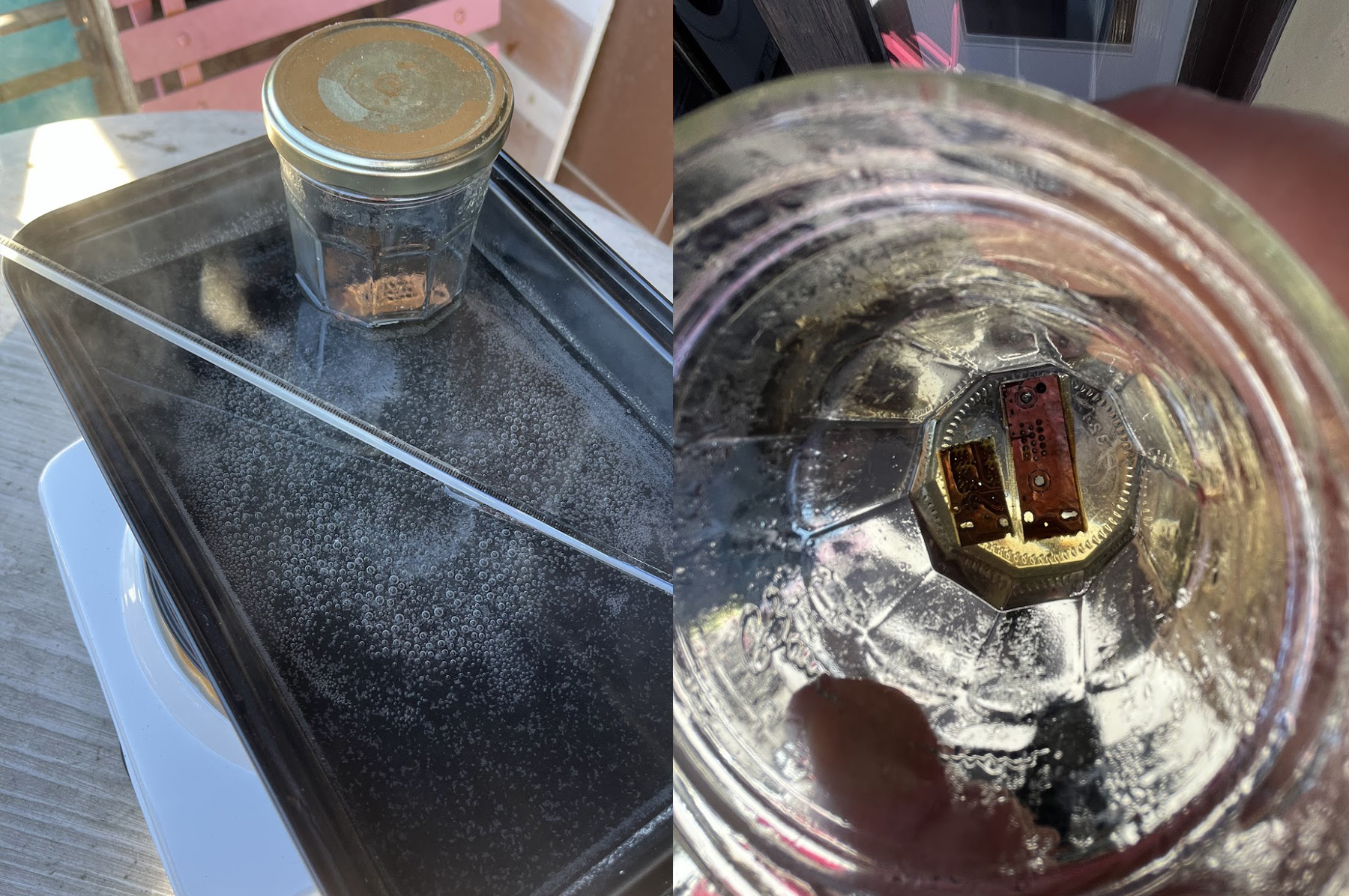}
  \caption{Use of DMSO for copper and resin recycling. Credit and source: Cyrus Khalatbari.}
  \label{fig:dmso}
\end{figure}

\begin{figure}[!htbp]
  \centering
  \includegraphics[width=\columnwidth]{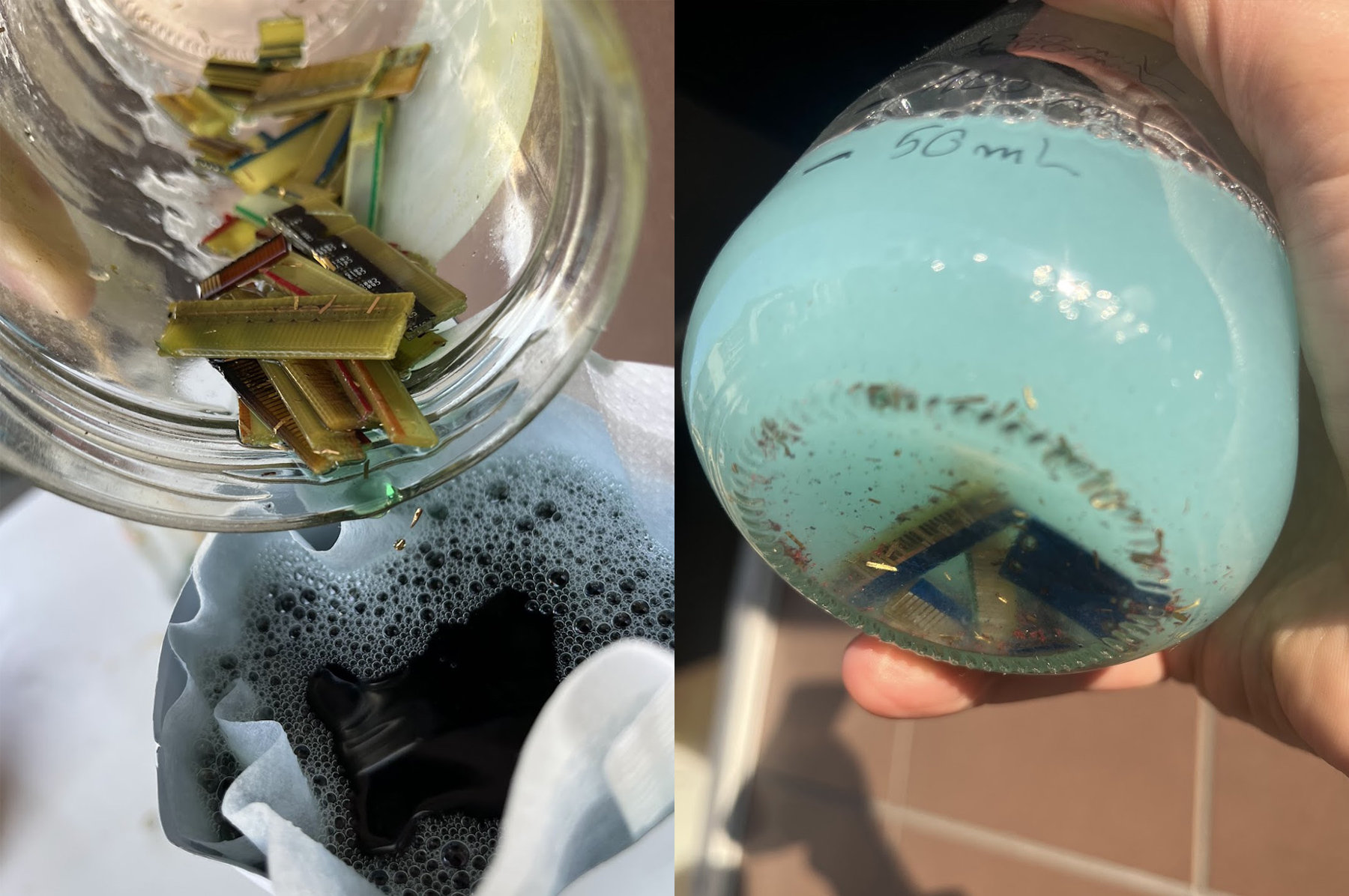}
  \caption{Gold extraction attempt. Credit and source: Cyrus Khalatbari.}
  \label{fig:gold}
\end{figure}

\textit{It all begins at Elliott's---in his apartment in Gaillard \footnote{Gaillard is a town located in Haute-Savoie, in the immediate vicinity of Geneva (Switzerland), where this research was conducted.}, on the two large tables of his workshop. Elliott, a chemist and electronics specialist I met in Geneva through a mutual friend, suggests we approach things methodically. We spread out all the acquired graphics cards and begin documenting them one by one, deciphering the references engraved on each circuit board and logging them in an Excel file. A simple Google search is enough to locate their models, retail prices, production years, and end-of-life dates. Dismantling begins with a screwdriver. From each card, we isolate three initial parts---the fan, the heat sink, the chassis---which we weigh before cataloguing. A heat gun then allows us to desolder the smaller components, collected in Ziploc bags [Fig.~\ref{fig:ziploc}]. This work of stripping eventually reveals the central piece on which everything rests: the printed circuit board. This resin substrate, threaded with copper connectors, is what structures and enables the card's functioning [Fig.~\ref{fig:circuitboard}]. To go further in analyzing the materials that compose it, we turn to chemistry---with a constraint we set ourselves from the outset: to prioritize low-tech techniques, reproducible in contexts like that of Ghana. The circuit board proves to be a stack of layers: a colored varnish on the surface, copper traces beneath, gold connectors along the edges which we cut with pliers, and a thick, non-flammable resin at the center. Elliott proposes a first entry point: a drain-unblocking liquid. Available in supermarkets, it nonetheless relies on caustic soda---NaOH---which, in a bain-marie maintained at 60°C for three hours, dissolves the varnish and brings the copper lines into detailed relief [Fig.~\ref{fig:caustic}]. To attack the central resin, we then turn to dimethyl sulfoxide---DMSO---present in certain medical products and orderable on Alibaba or Amazon. By immersing fragments of circuits from our cards in it, we observe, over 48 hours, the compact resin slowly decomposing into filaments. What was trapped is released: the copper layers can then be removed by hand [Fig.~\ref{fig:dmso}]. In a final session, we attempt to extract the gold from the cut connectors: alternating between hydrogen peroxide and vitamin C to precipitate the metal. After several days of tests and adjustments, we hit a limit our means cannot overcome. Unlike copper, iron, or aluminum---which the vendors at informal markets know how to extract with rudimentary tools---gold remains locked within a composite alloy [Fig.~\ref{fig:gold}]. Filtered, dissolved, purified multiple times, it remains elusive.}

\FloatBarrier

What conclusions can we draw from these explorations? Our gestures and techniques first confirmed what I had observed repeatedly in Ghana. On graphics cards, the materials recoverable through simple methods are those on the outside: aluminum, iron, copper. Massive, separable, accessible by hand. But what to do with the resin? How to interact with composite materials? Our experimentations progressively revealed the hidden face of miniaturization. The tantalum in capacitors, for example, is impossible to extract without industrial equipment. The smallest components are amalgams of metals layered in infinitesimally thin strata---there is little one can do with them. This is what McKenzie Wark describes as the ``mineral sandwich in your pocket''\cite{wark2019}: dozens of materials compressed into a volume so small that we lack the technology to disentangle them. Gold is the clearest example---without corrosive products and heavy infrastructure, it is unrecoverable. What these explorations ultimately brought to light is a dimension that dominant discourses on artificial intelligence tend to erase: the geological and elemental dimension of these infrastructures. Behind the algorithms lies matter---rare, difficult to extract, often impossible to recycle. Our hands on these circuits made that concrete, irrefutable.

\subsection{30 March, 14 April, 24 May 2025: Rebuild}

\begin{figure}[!htbp]
  \centering
  \includegraphics[width=\columnwidth]{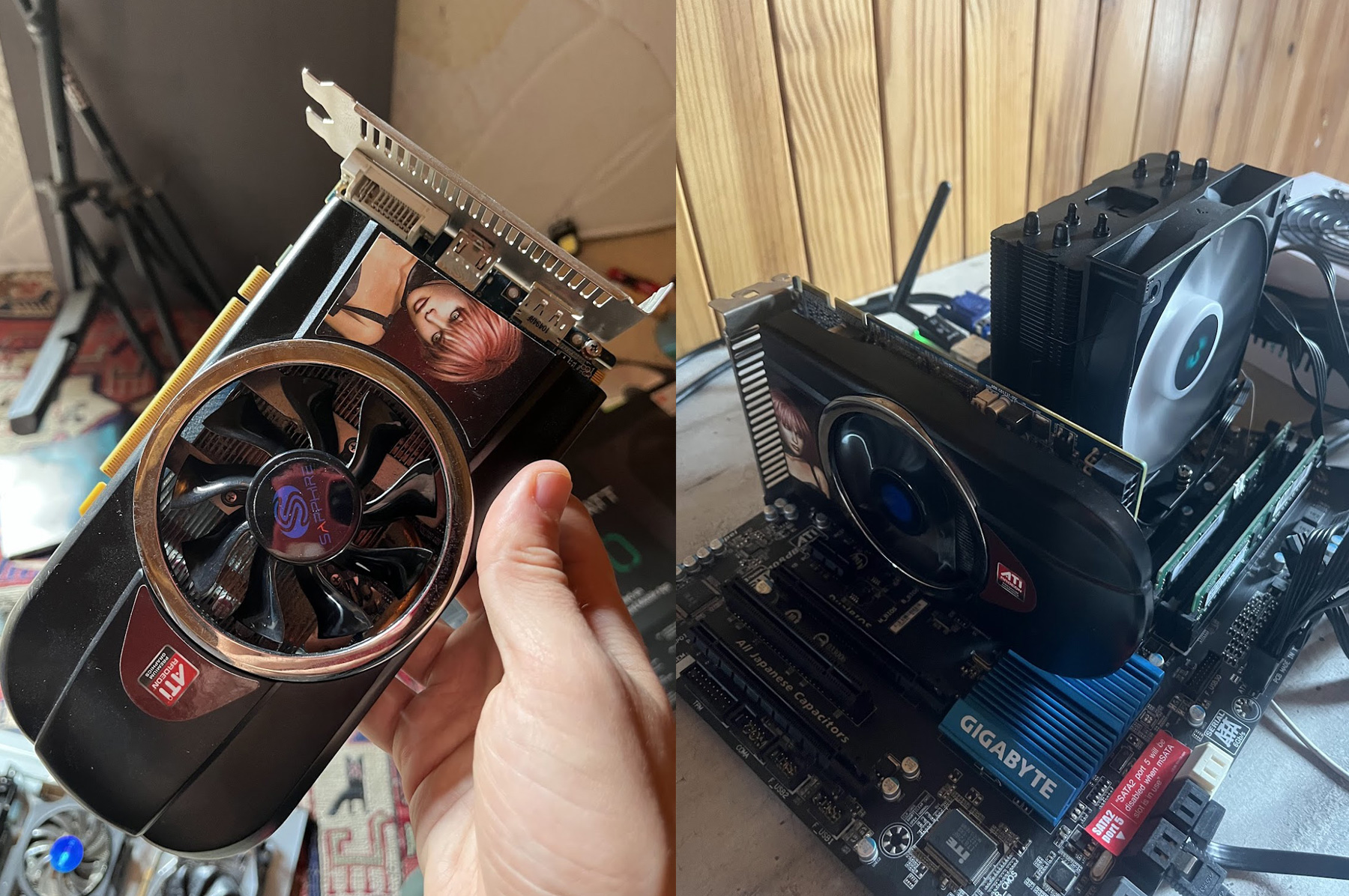}
  \caption{Sapphire Radeon HD 5770 1GB GDDR5 PCIe installed on its motherboard (right). Photography, Axel Azoulay and Cyrus Khalatbari, 2025}
  \label{fig:sapphire}
\end{figure}

\begin{figure}[!htbp]
  \centering
  \includegraphics[width=\columnwidth]{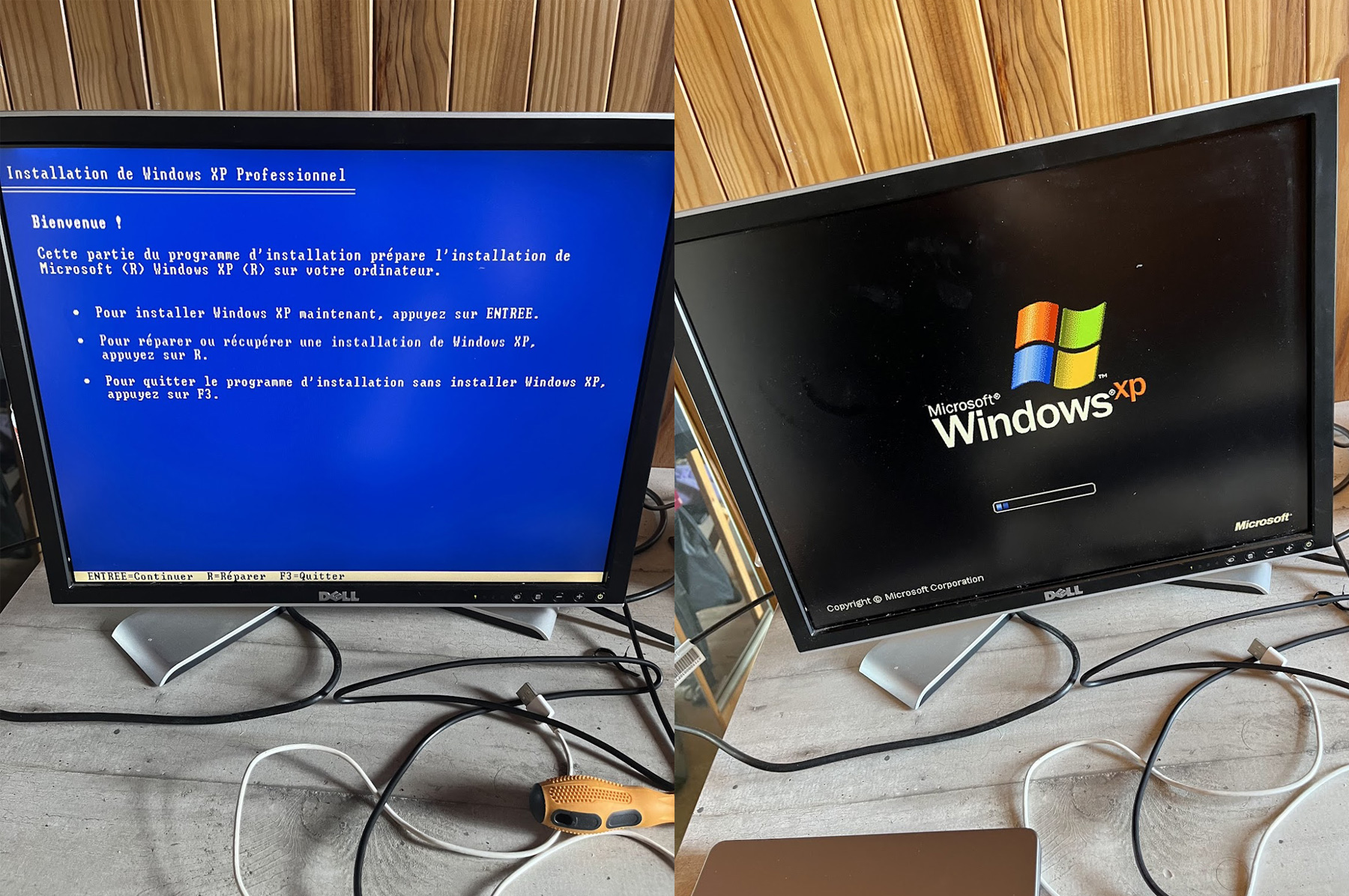}
  \caption{Installing Windows XP. Credit and source: Photography, Axel Azoulay and Cyrus Khalatbari, 2025}
  \label{fig:winxp}
\end{figure}

\begin{figure}[!htbp]
  \centering
  \includegraphics[width=\columnwidth]{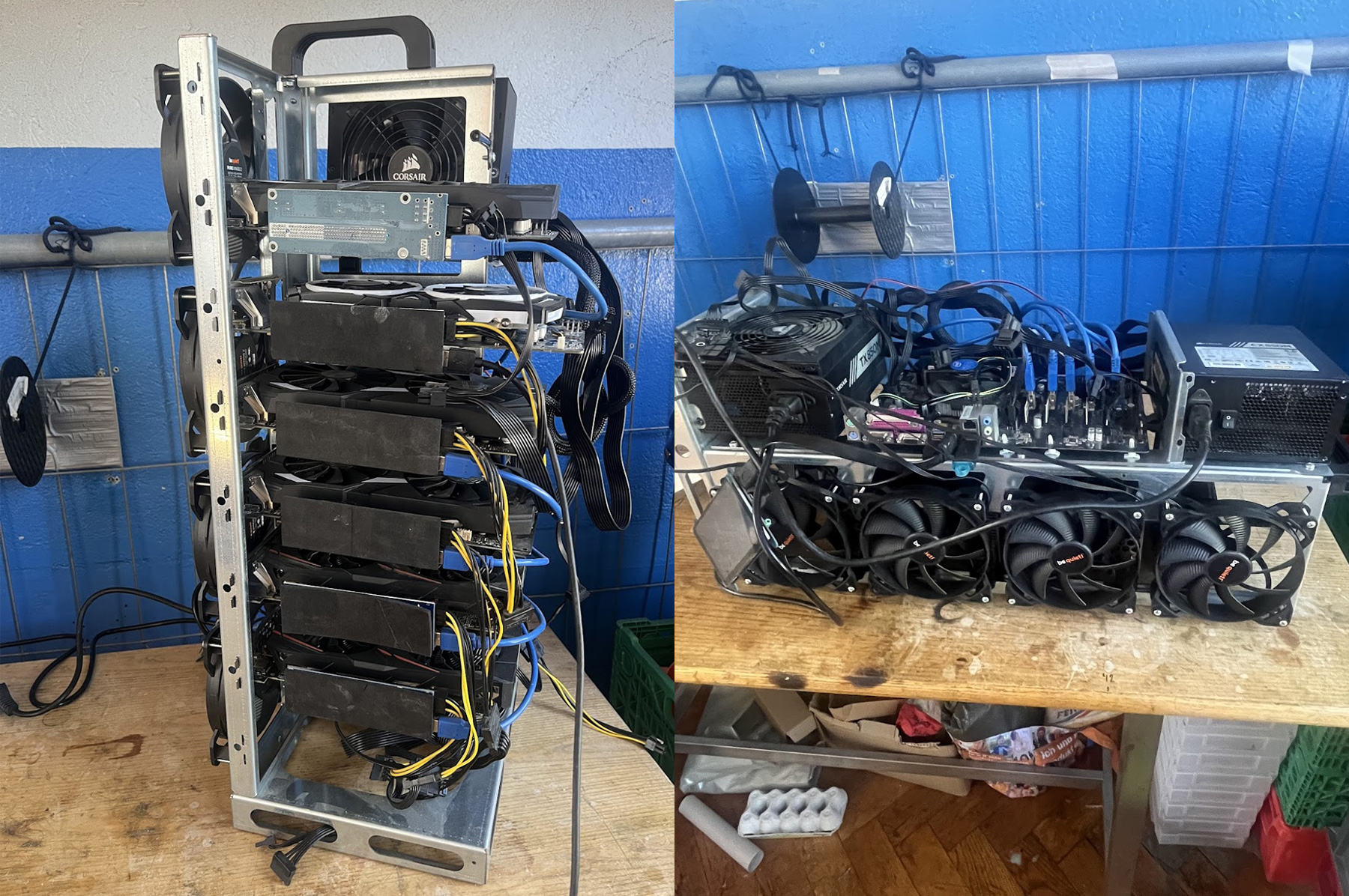}
  \caption{Acquired cryptocurrency mining rig. Credit and source: Cyrus Khalatbari.}
  \label{fig:miningrig}
\end{figure}

\begin{figure}[!htbp]
  \centering
  \includegraphics[width=\columnwidth]{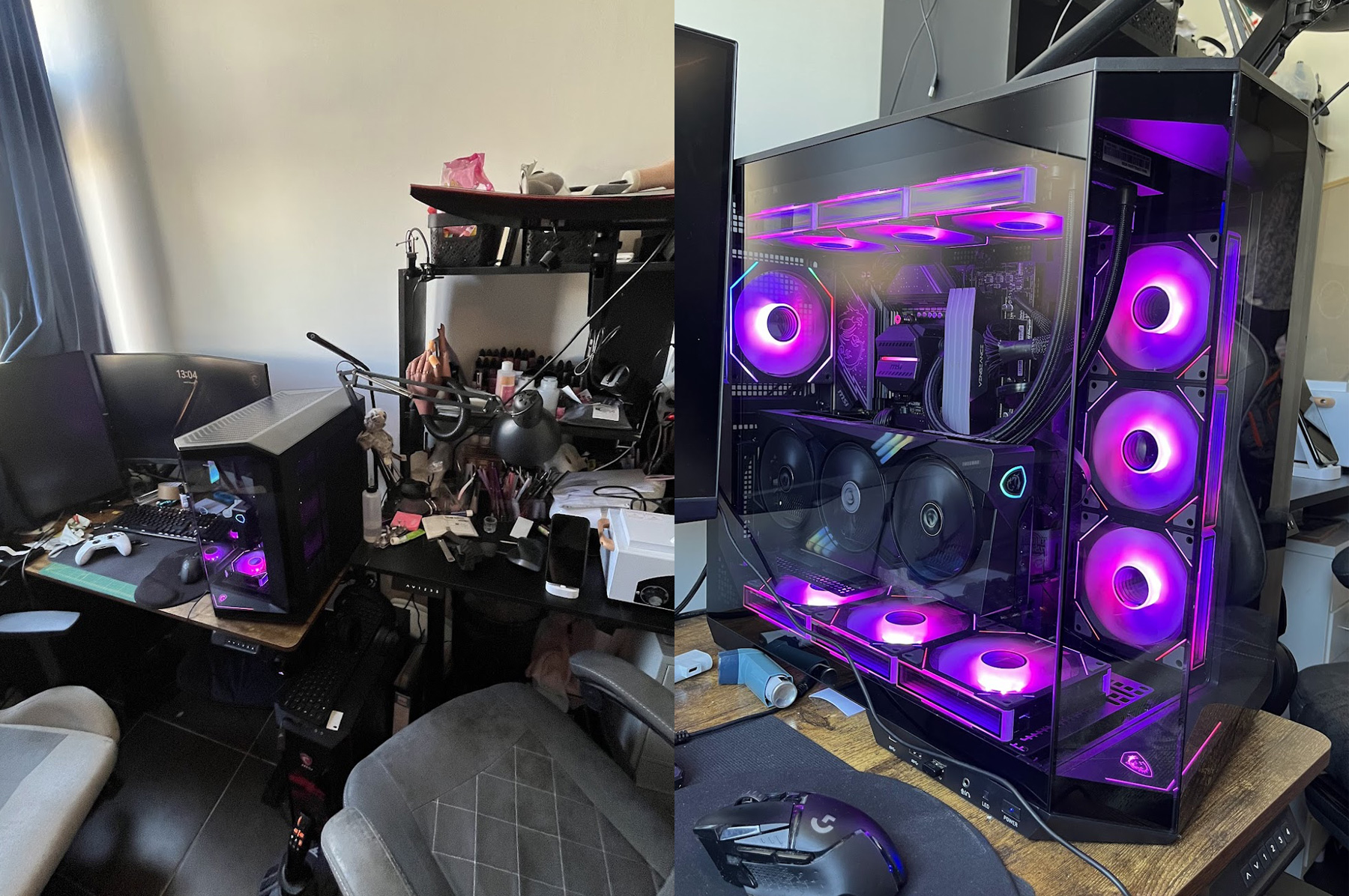}
  \caption{Enzo's gaming PC. Photography, Cyrus Khalatbari, 2025}
  \label{fig:gamingpc}
\end{figure}

\begin{figure}[!htbp]
  \centering
  \includegraphics[width=\columnwidth]{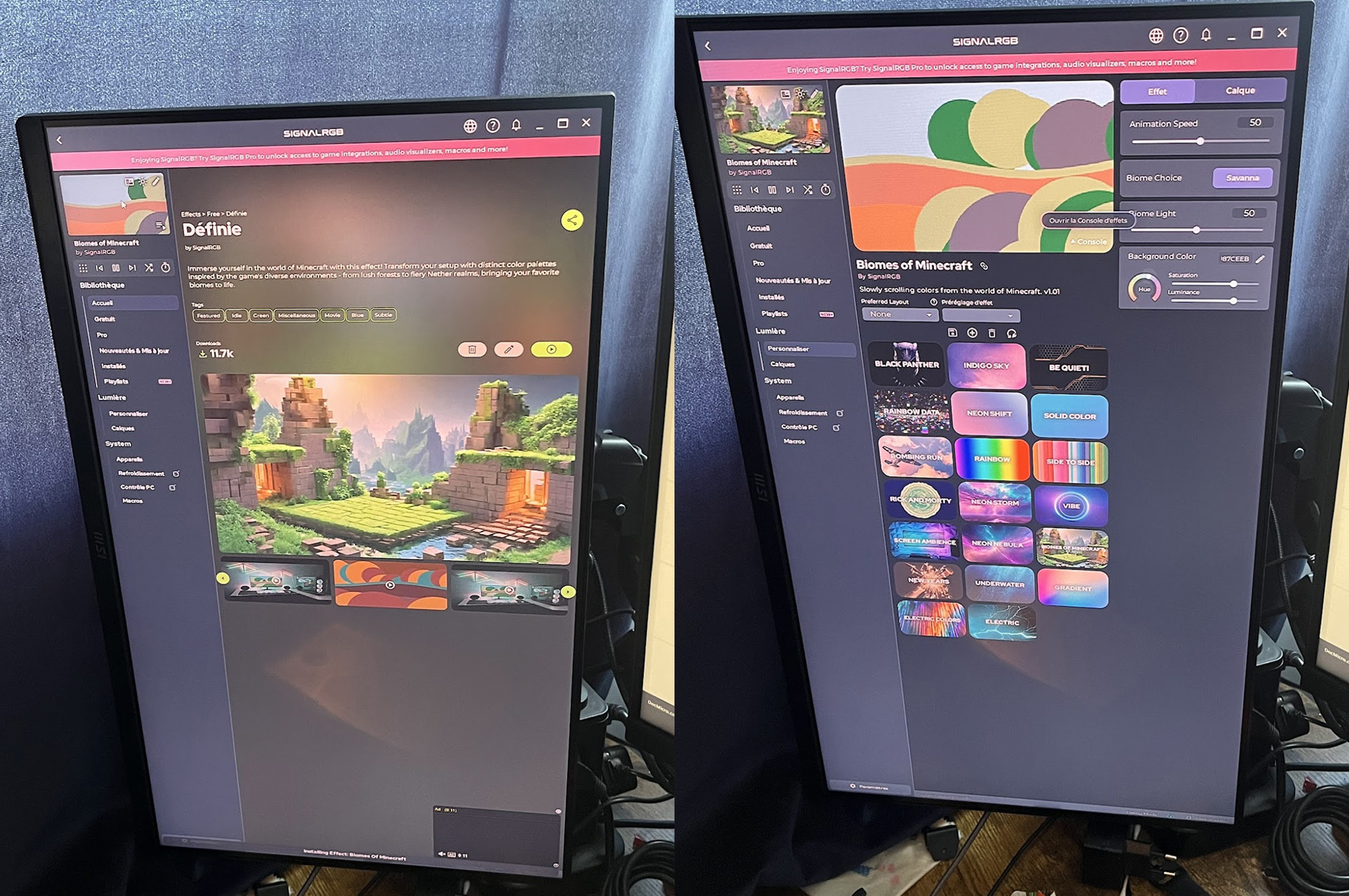}
  \caption{Various views of the SignalRGB interface. Photography, Cyrus Khalatbari, 2025}
  \label{fig:signalrgb}
\end{figure}

\textit{Alongside the activities with Elliott, I work with two other people from my circle: Axel and Enzo. Axel is a childhood friend. Passionate about video games from a young age, he quickly began building his own computers before becoming a programmer. I met Enzo, for his part, on Le Bon Coin, while procuring several electronic components and graphics cards. With Axel, we attempt to get the objects brought back from Ghana to function---a way of better understanding, through practice, the gestures of the maintenance technicians I encountered in the field. We choose a first card: the Sapphire Radeon HD 5770 1GB GDDR5 PCIe [Fig.~\ref{fig:sapphire}], brought to market in the late 2000s, purchased for 200 GHS from a reseller in Accra, and available on eBay for between 10 and 30 euros. We gather the necessary components---motherboard, memory sticks, connectors---then quickly hit a second obstacle: software dependencies. Axel decides: we need to install Windows XP [Fig.~\ref{fig:winxp}]. As the system starts up and its iconic wallpapers and animations come to life on screen, Axel points out something I would not have noticed on my own: all of this is running without our card being installed yet. It is the chipset managing it, he explains---before the GPU boom for AI, many computers simply didn't need one. Once the card is installed, we test it on the game Doom II \footnote{Doom II: Hell on Earth (1994) is a first-person shooter (FPS) developed by id Software, a direct sequel to Doom (1993). The player takes on the role of a marine fighting hordes of demons through labyrinthine environments.}. The results are striking: our game's gameplay becomes noticeably smoother. With Enzo, the experimentations take a different direction. Guided by what I observed in Taiwan, I acquire several recent cards, including two mining rigs each containing five Gigabyte Radeon RX Vega 56 cards. When they arrive at my workshop, Enzo is categorical: these rigs are not mounted efficiently [Fig.~\ref{fig:miningrig}]. He explains the concept of airflow \footnote{In the context of a cryptocurrency mining rig, airflow refers to the circulation of air through the rig, enabling the cooling of components (GPUs, motherboards) that generate large amounts of heat during mining.}: to expel the heat produced by mining, the cards must be oriented upward; otherwise, heat accumulates layer by layer. Screwdriver in hand, we rotate each card 90 degrees. A few days later, Enzo invites me to his home in Annemasse \footnote{Annemasse is a town located in Haute-Savoie, immediately adjacent to Geneva (Switzerland), where this research was conducted.}. His case takes up half his desk; an assembly of purple, yellow, and blue tones dynamically illuminates the internal components [Fig.~\ref{fig:gamingpc}]. This is SignalRGB \footnote{SignalRGB is a software that lets you control and synchronize the RGB lighting of all your computer components and peripherals (keyboards, mice, fans, graphics cards, etc.) from a single interface, while also offering the ability to create compositions that can be shared within its community.}, software enabling the downloading of colorimetric compositions produced by a community of users on Discord. Scrolling through the available creations, he stops on a grid of orange patches. I notice that some palettes push the concept very far---a grunge punk fluorescence that tips easily into a cyberpunk aesthetic. Combined with harmonious tones in the waterblocks, they make it possible to produce truly singular assemblages [Fig.~\ref{fig:signalrgb}].}

\FloatBarrier

What has this media archaeology taught us? First, to perceive the graphics card not as a fixed object, but as a node of entanglements. From geological strata to hardware components, all the way to operating systems, this insight became tangible with Axel when we realized that these cards brought back from Ghana, still in circulation, could only function with the installation of Windows XP---an operating system considered obsolete within our Western technological cultures. With Enzo, mounting these objects also allowed me to get as close as possible to the subcultures and communities of use that embody and articulate them. The colorimetric compositions of SignalRGB, the waterblocks, the airflow: this is where the thermoculture described by researcher Nicole Starosielski \cite{starosielski2016}---that attention to heat as matter to be managed, directed, but also exhibited, transforming thermal dissipation into an aesthetic gesture as much as a technical one---takes on its full meaning. In direct connection with the overclockers I met in Taiwan, these hybridizations thus allowed me to better situate, in other terms, the unifying, community-building, and meaning-making dimension of this object. In opposition to disembodied discourses about a global, abstract artificial intelligence, I was then better able to anchor, through these situated explorations, computing power and artificial intelligence within the social and cultural contexts that structure them.

\subsection{25 June, 7 July, 23 August 2025: Remix}

\begin{figure}[!htbp]
  \centering
  \includegraphics[width=\columnwidth]{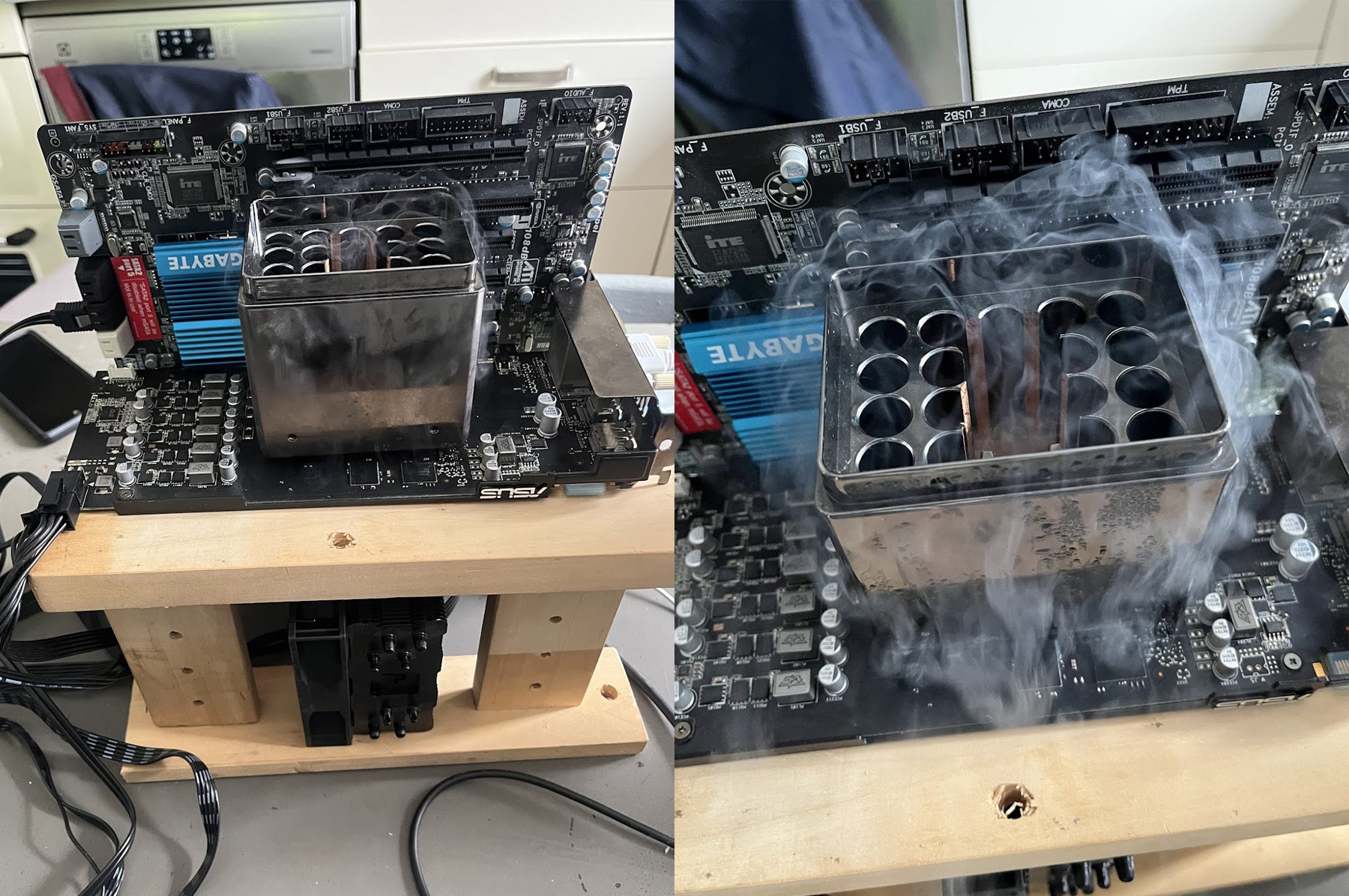}
  \caption{Alternative liquid nitrogen overclocking rig. Photography, Axel Azoulay and Cyrus Khalatbari, 2025}
  \label{fig:lnrig}
\end{figure}

\begin{figure}[!htbp]
  \centering
  \includegraphics[width=\columnwidth]{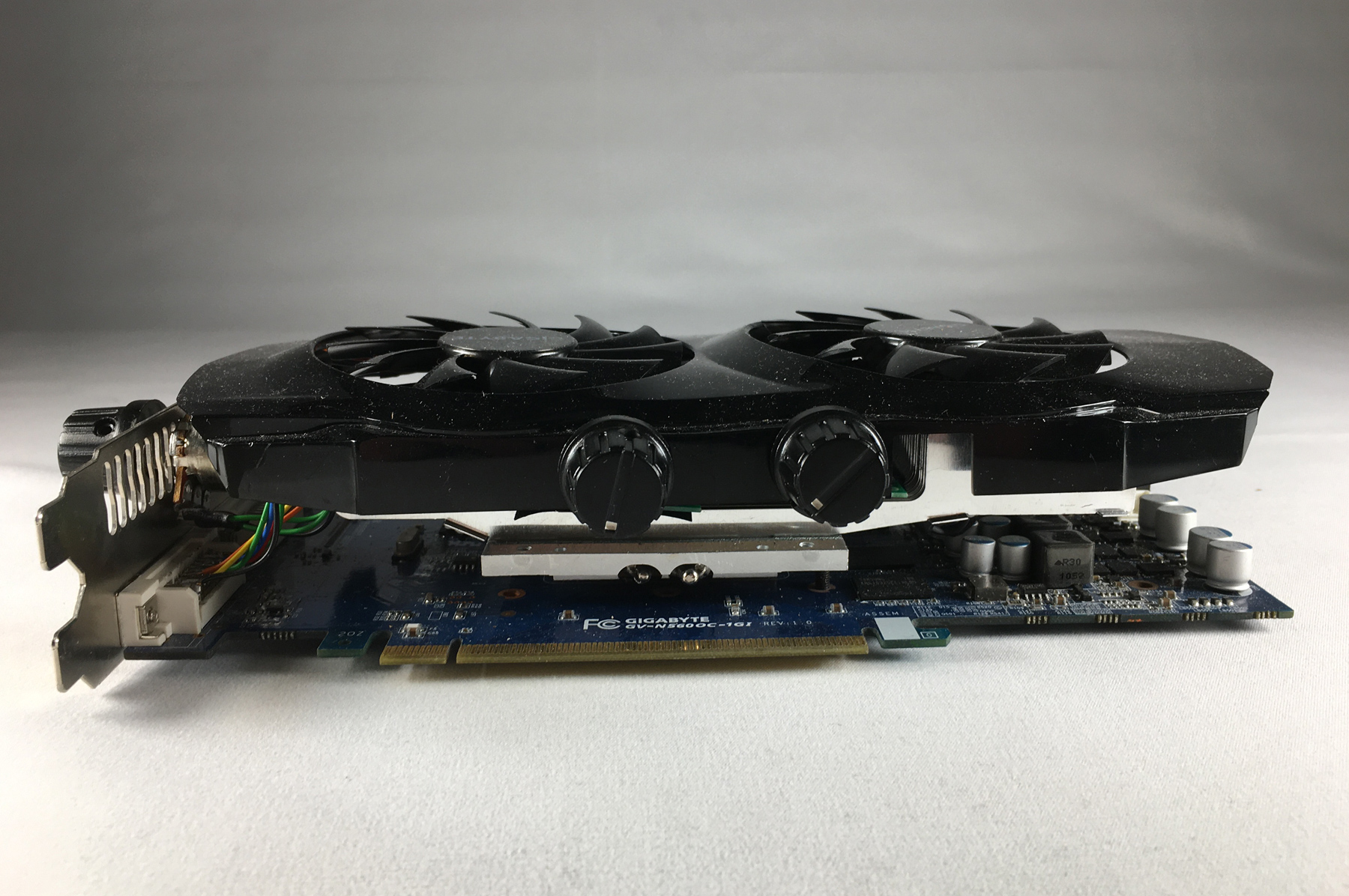}
  \caption{Prototype: Synth Fan. Photography, Elliott Croset and Cyrus Khalatbari, 2025}
  \label{fig:synthfan}
\end{figure}

\begin{figure}[!htbp]
  \centering
  \includegraphics[width=\columnwidth]{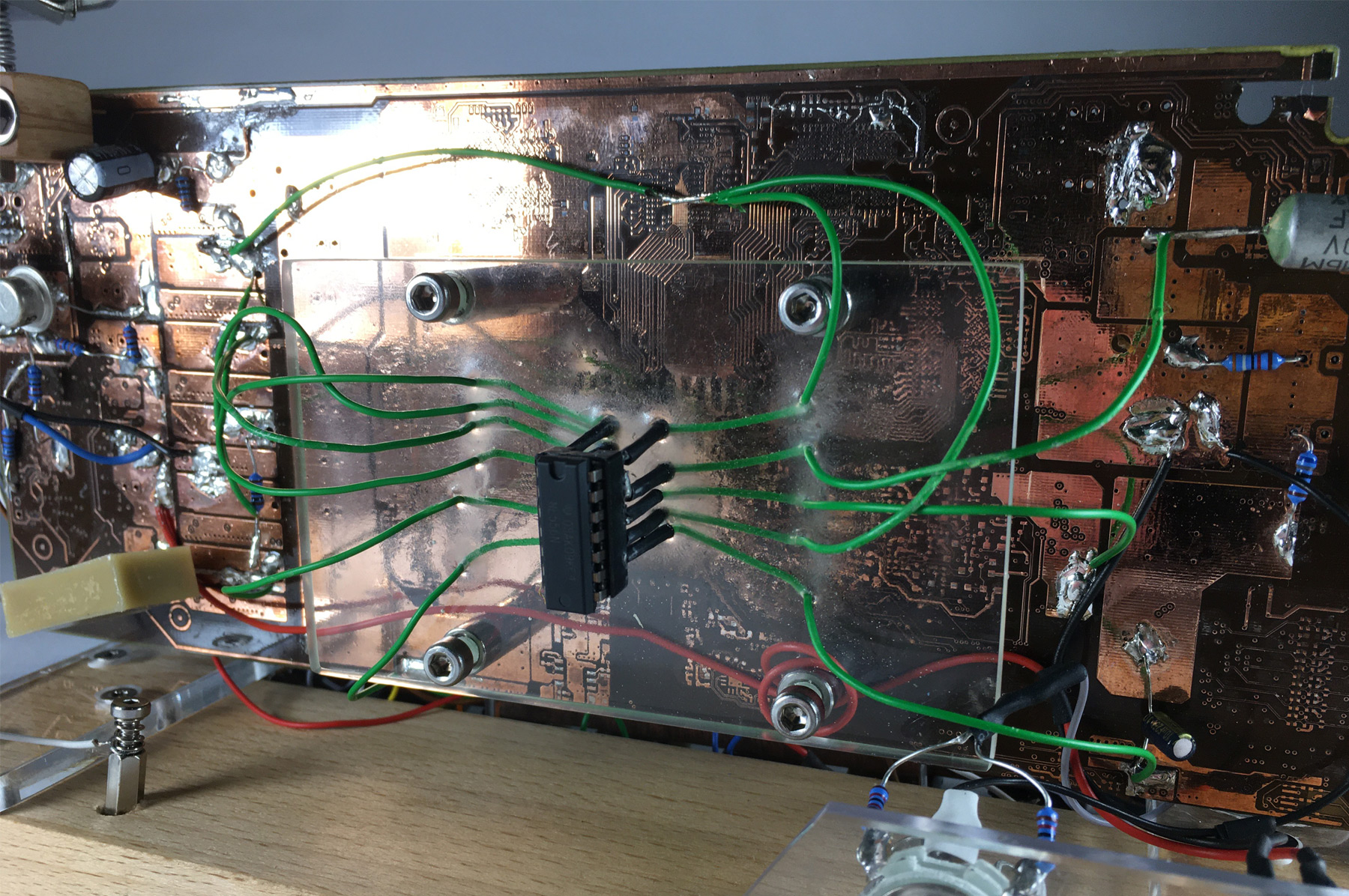}
  \caption{Prototype: APC. Photography, Elliott Croset and Cyrus Khalatbari, 2025}
  \label{fig:apc}
\end{figure}

\begin{figure}[!htbp]
  \centering
  \includegraphics[width=\columnwidth]{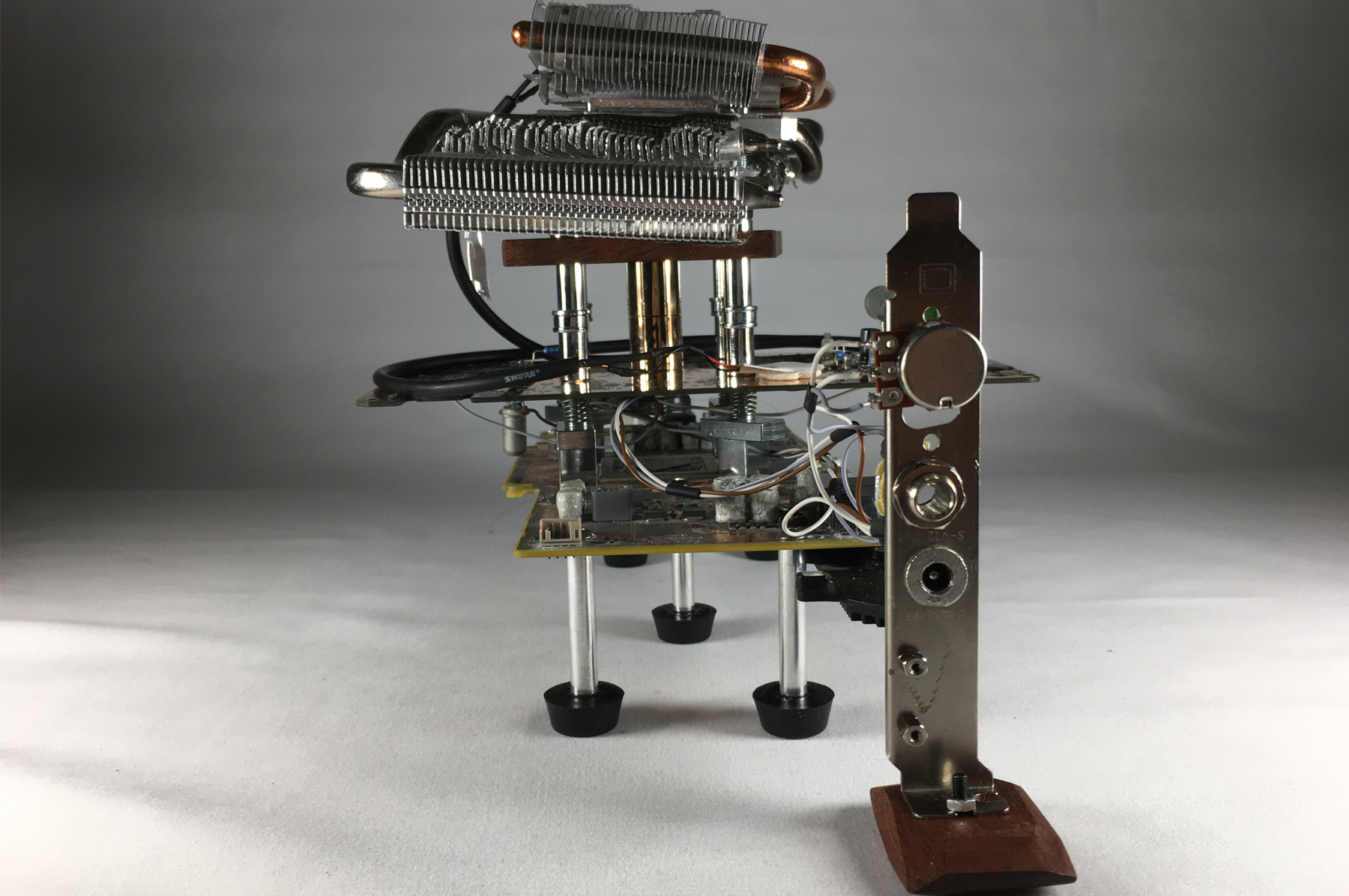}
  \caption{Prototype: Piezo noisebox. Photography, Elliott Croset and Cyrus Khalatbari, 2025}
  \label{fig:piezo}
\end{figure}

\begin{figure}[!htbp]
  \centering
  \includegraphics[width=\columnwidth]{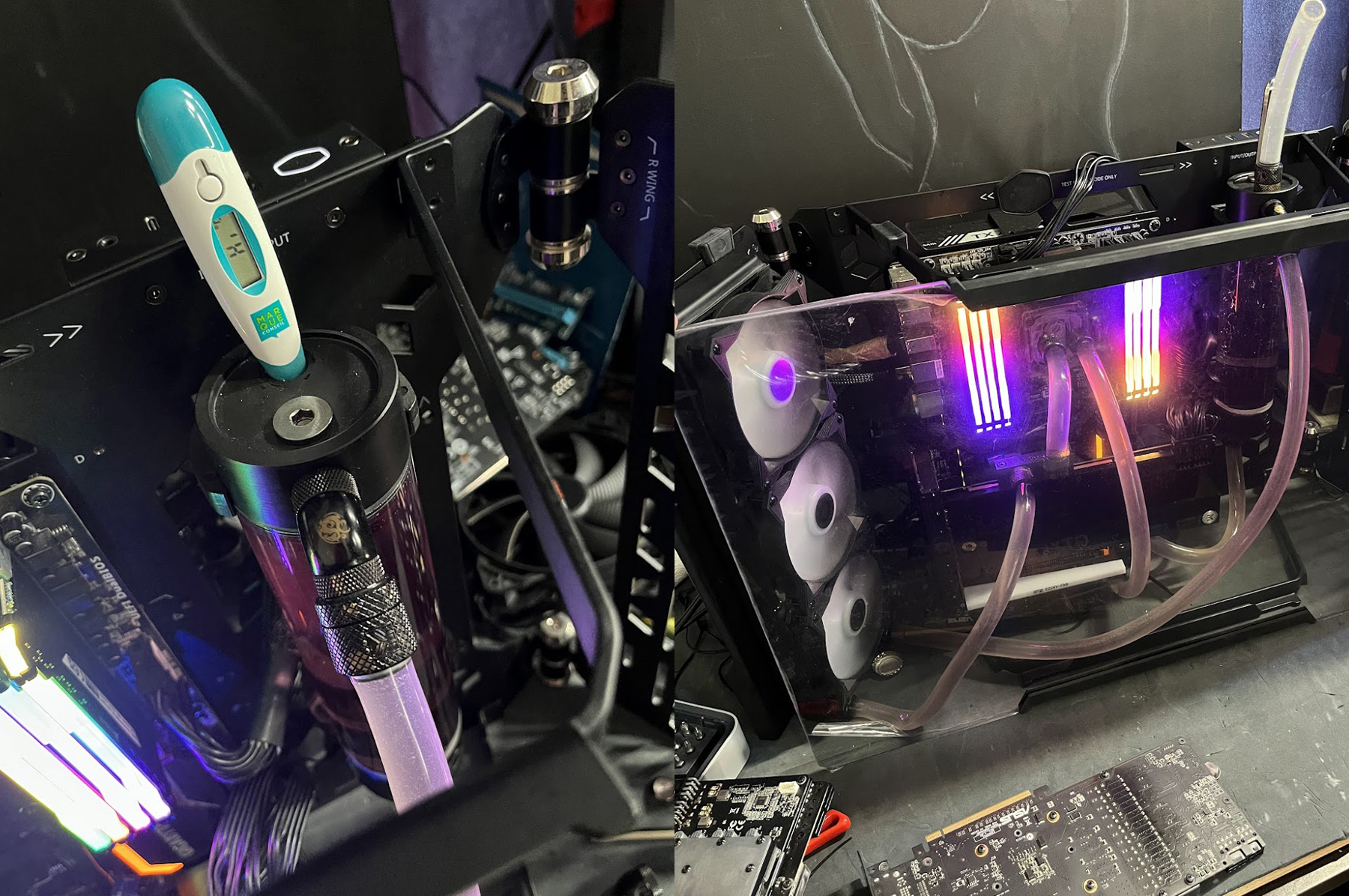}
  \caption{Prototype: PC $\times$ Mod $\times$ Recycling. Photography, Elliott Croset, Enzo Quebrillac and Cyrus Khalatbari, 2025}
  \label{fig:pcmod}
\end{figure}

\textit{As my conversations with Elliott, Axel, and Enzo progressed, a third type of exploration imposed itself: making my two fieldsites in Taiwan and Ghana speak to each other through the creation of deliberately anachronistic objects and gestures, hybridizing heterogeneous materialities and temporalities---obsolete cards and cutting-edge techniques, performance logics and scrap logics, gaming cultures and recycling practices. A first thread emerged with Axel around liquid nitrogen. Rather than using it on the latest-generation cards as in Taiwan, we chose to test its effect on our Sapphire Radeon acquired in Ghana---a deliberately displaced gesture, applying an extreme overclocking technique to an object considered end-of-life. We began by constructing an overclocking platform: by tilting the motherboard 90 degrees, a small wooden structure enabled the GPU to be positioned horizontally to pour nitrogen onto it [Fig.~\ref{fig:lnrig}]. Once the cooler was removed, a first problem arose: the heat generated at system startup forced the card to cut out automatically. It was then that the role of cooling appeared in all its brutality---only a continuous flow of nitrogen, evaporating instantly on the silicon, made the slightest interaction with the system possible. The repeated failure of these operations---several broken cards---opened a second thread, oriented this time toward low-tech artistic reappropriation. This was also motivated by the need to develop platforms for exchange and collaboration through making with the Ghanaian communities connected to my work---a way of building spaces for shared practice rather than simple relations of observation or collection. Our conversations with Elliott led us to work with the waste itself: circuit boards, fans, heat sinks. The idea was to produce small synthesizers drawing on the material specificities of each card, transforming objects destined for the landfill into experimental musical instruments. By controlling each fan through electrical pulse and placing a light sensor beneath the component, we created a first instrument translating luminosity values into sound frequencies [Fig.~\ref{fig:synthfan}]. A second exploration focused on the copper connectors of the circuits: by soldering directly from these traces, we recreated the Atari Punk Console, a well-known synthesizer in the world of experimental music [Fig.~\ref{fig:apc}]. This series concluded with a third prototype, a piezo noisebox built from the heat sink, where the link between the material specificities of each GPU and the sonic possibilities appeared in particularly explicit form: the shape of each heat sink directly influencing the resonances and timbres produced [Fig.~\ref{fig:piezo}]. From these sonic objects, a third thread naturally followed, this time focused on cards still in operation. Drawing inspiration from my Taiwanese fieldwork, I opted for the creation of PC mods in which the heat produced by the GPUs would be used actively rather than simply dissipated. With Elliott and Enzo, we installed water cooling systems on several cards. After a few temperature readings, an idea imposed itself: to use this residual heat to create a bain-marie enabling the progressive thermal recycling of card waste---resin, copper---joining back up with our first experimentations in the field [Fig.~\ref{fig:pcmod}]. In doing so, my two fieldsites finally met, putting into dialogue through this object two sociotechnical processes---production and waste, Taiwan and Ghana---so often perceived as opposites.}

\FloatBarrier

What can we take away from these d\'{e}tournements? They trace, at bottom, a single shared preoccupation: breaking the linear temporality that our Western cultures project onto information technologies. As digital culture researcher Tung-Hui Hu shows in \textit{Prehistory of the Cloud} \cite{hu2015}, our digital infrastructures are built on colonial inheritances and Cold War logics that dominant discourse carefully erases. By applying liquid nitrogen to an end-of-life card, by fabricating synthesizers from waste, by using a GPU's heat to melt its own matter, these projects present computing power not as a technological rupture but as a genealogy of entangled temporalities---what designer Garnet Hertz and media theorist Jussi Parikka call ``zombie'' media \cite{hertz2012}: forgotten or marginalized techniques summoned to rethink the dominant narratives of our digital culture. These explorations also sought to displace another binary: the one that separates users and workers, technology and labor. In the spirit of what researcher Sasha Costanza-Chock \cite{costanza-chock2020} formulates with the principle ``Nothing about us without us,'' the aim was not to produce objects disconnected from the field, but to develop workshops and spaces of shared fabrication with the communities encountered in Ghana---thereby proposing alternatives to purely transactional gestures. By simultaneously weaving collaborations with engineers and chemists such as Axel, Enzo, and Elliott, these projects also made it possible to build a concrete sociotechnical literacy: a way of situating our objects, of nuancing the disembodied discourses on artificial intelligence, and of making sensible again what miniaturization conceals---the matter, the gesture, the labor.

\FloatBarrier

\section{Conclusion: Toward New Literacies through Making}

This field notebook has been, above all, a space of demystification through making. By dismantling, reconstituting, and repurposing some fifty graphics cards acquired in Ghana and Taiwan, I sought to make tangible the realities that miniaturization methodically conceals. For it is precisely this material compression that feeds the grand, disembodied narratives of artificial intelligence---that of a technology that is abstract, apolitical, freed from all geography and all labor. The more the components that underpin it shrink and grow opaque, the easier it becomes to ignore the rare metals they contain, the workers who assemble them, the communities who inherit their waste. These situated explorations allowed me to weave connections between worlds that these narratives deliberately keep apart -- performance and scrap, Taipei and Accra, gaming, and informal recycling. As I have demonstrated through this investigative and analytical notebook, research-creation methods---dismantling, reconstruction, remix---constitute full epistemologies in their own right for apprehending what seems to us a priori external, opaque, or inaccessible. By laying hands on these circuits, by grinding and resoldering them otherwise, I accessed dimensions that neither distant observation nor textual analysis could have revealed. The technical gesture thus becomes a critical gesture: a way of reopening the black box, of restoring artificial intelligence to its tangible materialities, and of collectively proposing other sociotechnical imaginaries. In doing so, this work contributes to the field of ICT for sustainability, affirming research-creation as a rigorous means of disentangling the material and environmental realities that computational systems tend to render invisible.



\onecolumn

\end{document}